\documentclass[]{elsarticle}
\makeatletter
\def\ps@pprintTitle{%
	\let\@oddhead\@empty
	\let\@evenhead\@empty
	\def\@oddfoot{}%
	\let\@evenfoot\@oddfoot}
\makeatother
\usepackage{fancyhdr}
\usepackage{epsfig}
\usepackage{amsfonts}
\usepackage{amssymb, graphics, amsmath}
\usepackage{dsfont}
\usepackage{hyperref}
\usepackage{color}
\usepackage{pdflscape}
\usepackage{pifont}
\usepackage{graphicx} 
\usepackage{bm}
\usepackage{slashed}
\usepackage[compat=1.1.0]{tikz-feynman}
\usepackage{tikz}
\usepackage[compat=1.1.0]{tikz-feynhand}

\allowdisplaybreaks

\def\slashchar#1{\setbox0=\hbox{$#1$}           % set a box for#1
	\dimen0=\wd0                                    % and get its size
	\setbox1=\hbox{/} \dimen1=\wd1                  % get size of/
	\ifdim\dimen0>\dimen1                           % #1 is bigger
	\rlap{\hbox to \dimen0{\hfil/\hfil}}            % so center / in box
	#1                                             % and print #1
	\else                                          % / is bigger
	\rlap{\hbox to \dimen1{\hfil$#1$\hfil}}        % so center #1
	/                                           % and print/
	\fi}       

\begin{document}
\title{Extracting the low-energy constant $L_0^r$ at three flavors\\ from pion-kaon scattering}

\author[bonn]{Chaitra Kalmahalli Guruswamy }
\author[bonn,fzj,tbilisi]{Ulf-G. Mei\ss{}ner}
\author[bonn]{Chien-Yeah Seng}

\address[bonn]{Helmholtz-Institut f\"ur Strahlen- und
	Kernphysik and Bethe Center for Theoretical Physics,\\
	Universit\"at Bonn, D-53115 Bonn, Germany}
\address[fzj]{Institute for Advanced Simulation, Institut f\"ur Kernphysik and
	J\"ulich Center for Hadron Physics,\\ Forschungszentrum J\"ulich,
	D-52425 J\"ulich, Germany}
\address[tbilisi]{Tbilisi State University, 0186 Tbilisi, Georgia}

\begin{abstract}
Based on our analysis of the contributions from the connected and disconnected contraction
diagrams to the pion-kaon scattering amplitude, we provide the first determination of the
only free low-energy constant at $\mathcal{O}(p^4)$, known as $L_0^r$, in  SU$(4|1)$
Partially-Quenched Chiral Perturbation Theory using the data from the
Extended Twisted Mass Collaboration, $L_0^r(\mu=M_\rho) = 0.77(20)(25)(7)(7)(2)\cdot 10^{-3}$.
The theory uncertainties originate from the unphysical scattering length, the physical
low-energy constants, the higher-oder chiral corrections, the (lattice) meson masses and
the pion decay constant, respectively.  
\end{abstract}

\maketitle

\thispagestyle{fancy}

\section{Introduction}

Partially-Quenched Quantum Chromodynamics (PQQCD) and its low-energy effective field theory (EFT),
Partially-Quenched Chiral Perturbation Theory(PQChPT)~\cite{Bernard:1993sv,Sharpe:1999kj,Sharpe:2000bc,Sharpe:2001fh,Bernard:2013kwa,Sharpe:2006pu,Golterman:2009kw}, are powerful tools to assist the
first-principles calculations of hadronic observables using lattice QCD. They were originally
created to handle the so-called partially-quenched approximation in early lattice studies,
where the ``valence'' and ``sea'' quark masses were made distinct in order to simplify the
calculation of the fermion determinant. Nowadays such an approximation has been largely abandoned,
but the devised theory frameworks have found their own ways to continue being useful.
In particular, it was recently realized that since additional quark flavors are introduced in PQChPT, it
allows an EFT description of each individual quark contraction diagram in the lattice simulation of
a given physical observable, which is very useful for getting a better handle on
the noisier and computationally-expensive contractions (the so-called ``disconnected diagrams'').
This idea was applied initially to the study of the hadronic vacuum
polarization~\cite{DellaMorte:2010aq} and the pion scalar form factor~\cite{Juttner:2011ur},
and was later extended to pion-pion scattering~\cite{Acharya:2017zje,Acharya:2019meo} and
the parity-odd pion-nucleon coupling constant~\cite{Guo:2018aiq}.

As in any EFT, the full predictive power of PQChPT to a given order is guaranteed only when
all the low-energy constants (LECs) at that order are fixed. This is a non-trivial task since
some of them are not constrained by any physical experiment and can only be determined through
lattice simulations. In particular, in an extended flavor sector with $N_f\geq 4$, the
Cayley-Hamilton relation is no longer valid, which leads to the following term in the PQChPT
Lagrangian at $\mathcal{O}(p^4)$:
\begin{equation}
  {\cal L}^{(4)} = L_0 \, {\rm Str} \left(\partial_\mu U^\dagger \partial_\nu U
  \partial^\mu U^\dagger \partial^\nu U\right)~,
\end{equation}
where `Str' denotes the supertrace over the extended flavor space, $U$ is the standard
exponential representation of the pseudo-Nambu-Goldstone bosons, and $L_0$ is a new LEC that
does not appear in ordinary two-flavor and three-flavor ChPT. Despite appearing at 
next-to-leading order (NLO), this LEC contributes to static quantities such as the pion mass
and decay constant only at the next-to-next-to-leading order (NNLO), when the valence and
sea quark masses are kept distinct. Based on this, Ref.~\cite{Boyle:2015exm} has determined
the renormalized LEC $L_0^r$ in the SU$(4|2)$ PQChPT, which is equivalent to a two-flavor ChPT
in computations of physical processes. The quoted result is $L_0^r(\mu=1~\mathrm{GeV}) =
1.0(1.1)\cdot 10^{-3}$, using dimensional regularization.

Refs.~\cite{Acharya:2017zje,Acharya:2019meo} provided an alternative determination of $L_0^r$
in SU(4$|$2) using the fact that it appears in separate contraction diagrams in the $\pi\pi$
scattering amplitude at NLO, and therefore can be obtained from the unphysical scattering length
defined through an appropriate linear combination of contraction diagrams. The advantage of
this new method is that, since only light quarks are involved in the procedure, the higher-order
chiral corrections that scale generically as $M_\pi^2/(4\pi F_\pi)^2$ are expected to be small
so the NLO fitting is more stable than NNLO, and of course the number of unknown LECs in the
former is also smaller. Using the lattice data of connected $\pi\pi$ correlation functions by
the Extended Twisted Mass (ETM) Collaboration~\cite{Helmes:2015gla}, Ref.~\cite{Acharya:2019meo}
reported $L_0^r(\mu=1~\mathrm{GeV}) = 5.7(1.9)\cdot 10^{-3}$. The significant difference with the
result from the NNLO fit in Ref.~\cite{Boyle:2015exm} is yet to be understood. Resolving such
a discrepancy may improve our knowledge of the chiral dynamics at low energies, or
even reveal some unexpected lattice systematics that could play important roles in precision physics. 

The analysis in Refs.~\cite{Acharya:2017zje,Acharya:2019meo} could be straightforwardly generalized
to the three-flavor case. In particular, the so-called SU(4$|$1) PQChPT is of special interest
because it is the simplest extension of the original three-flavor ChPT. Moreover, among all its
renormalized LECs at $\mathcal{O}(p^4)$, only $L_0^r$ is undetermined, while all the others
are identical to those in SU(3). This implies that the theory would be fully predictive at
$\mathcal{O}(p^4)$ once $L_0^r$ is fixed, and could then be used to aid the lattice studies of
interesting hadronic processes involving strange quarks, such as $K\pi\to K\pi$, $\pi\pi\to K\bar{K}$
and $K\eta\to K\eta$ scatterings, in particular channels where disconnected diagrams appear
(e.g. the $I=1/2$ channel of $K\pi$ scattering).  

In this letter, we present the first-ever numerical determination of $L_0^r$ in SU(4$|$1) 
based on the method outline in our previous paper, Ref.~\cite{Guruswamy:2020uif}. The main
idea is that, by switching the relative sign between the two types of connected contraction
diagrams that occur in the $I=3/2$ $K\pi$ scattering, one obtains effectively an unphysical
single-channel scattering amplitude $T_\beta$, of which the scattering length depends on $L_0^r$
at $\mathcal{O}(p^4)$. Invoking the recent lattice data by the
ETM collaboration~\cite{Helmes:2018nug}, the unphysical scattering length is obtained through
the usual L\"{u}scher analysis of the discrete energy states~\cite{Luscher:1986pf}, which
consequently fixes $L_0^r$. This completes the SU(4$|1$) PQChPT Lagrangian at $\mathcal{O}(p^4)$
and is potentially useful for all the purposes mentioned above. Besides, it serves as a prototype
for future analyses of  more complicated PQ-extensions of three-flavor ChPT. Such theories, after
integrating out the strange quark, reduce to PQChPT with only light flavors. This may give
an independent check of value of $L_0^r$ in SU(4$|$2), and provide some hints towards the
solution to the aforementioned discrepancy.

\section{Formalism}

\begin{figure}[t!]
	\centering
	\includegraphics[width=0.2\linewidth]{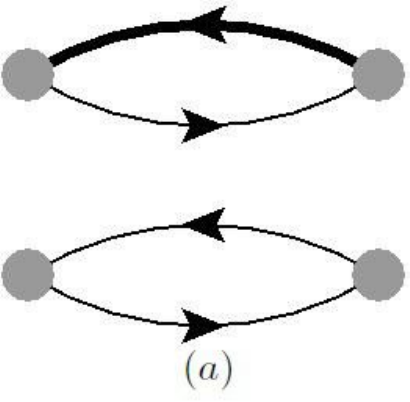}\qquad\qquad\includegraphics[width=0.2\linewidth]{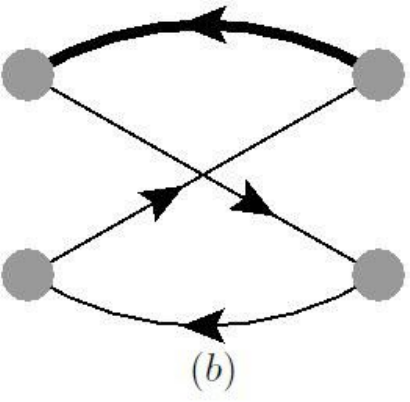}
	\caption{\label{fig:TaTb}Connected contraction diagrams $T_a$ and $T_b$. The thick line indicates an $\langle s\bar{s}\rangle$ contraction.}
\end{figure}

As detailed in Ref.~\cite{Guruswamy:2020uif}, the pertinent two contractions in $K\pi$
scattering with total isospin $I=3/2$, as depicted in Fig.\ref{fig:TaTb}, are related to the
amplitudes in the SU$(4|1)$ PQChPT:
\begin{eqnarray}
T_{a}(s,t,u) & = & T_{(u\bar{s})(d\bar{j})\rightarrow(u\bar{s})(d\bar{j})}(s,t,u)~, \nonumber \\
T_{b}(s,t,u) & = & T_{(u\bar{s})(d\bar{j})\rightarrow(d\bar{s})(u\bar{j})}(s,t,u)~,\label{eq:TaTb}
\end{eqnarray}
with $s,t,u$ the conventional Mandelstam variables subject to the constraint $s+t+u =
2(M_K^2+M_\pi^2)$, $u,d,s$ are the physical quarks and $j$ denotes the additional
valence quark (which comes together with a ghost quark $\tilde{j}$) in PQChPT. The assigned
quark masses are $m_u=m_d=m_j=m_{\tilde{j}}<m_s$. For total isospin $I=1/2$,
there is one additional scattering amplitude $T_c$ that is related to $T_b$ through a simple
crossing, $T_c(s,t,u)\equiv T_b(u,t,s)$. Although such a crossing is analytically straightforward,
from a lattice point of view $T_a$ and $T_b$ are relatively easy to evaluate, whereas $T_c$
involves a pair of quark propagators that start and end on the same temporal slice, and is
exactly what we call a ``disconnected diagram''. Such diagram suffers from a low signal-to-noise
ratio, and represents a fundamental challenge in the first-principles study of the $K\pi$
scattering in the $I=1/2$ channel. 

From the above one may define three effective single-channel scattering amplitudes:
\begin{eqnarray}
 T_{\alpha}(s,t,u)&=&T_a(s,t,u)+T_b(s,t,u)\nonumber\\
T_{\beta}(s,t,u)&=&T_a(s,t,u)-T_b(s,t,u)\nonumber\\
T_{\gamma}(s,t,u)&=&T_{a}(s,t,u)-\frac{1}{2}T_b(s,t,u)+\frac{3}{2}T_c(s,t,u)~,
\end{eqnarray} 
where $T_\alpha$ and $T_\gamma$ correspond to the physical $I=3/2$ and $I=1/2$ scattering amplitudes,
respectively, while $T_\beta$ is an unphysical amplitude. The S-wave scattering lengths are
defined through the threshold values of the amplitudes:
\begin{equation}
a_0^i = -\frac{1}{8\pi\sqrt{s_0}}T_i(s_0,t_0,u_0)\quad (i=\alpha,\beta,\gamma)~,
\end{equation}
with $s_0 = (M_K+M_\pi)^2$, $t_0=0$ and $u_0=(M_K-M_\pi)^2$. Obviously, only the unphysical scattering
length $a_0^\beta$ can depend on the unphysical LEC $L_0^r$. Its  explicit expression at
$\mathcal{O}(p^4)$ reads:
\begin{eqnarray}
a_{0}^{\beta} & = & \frac{-1}{8 \pi\sqrt{s_{0}} }\biggl[\frac{M_{\pi }^2 M_K^2}{F_{\pi }^4}\biggl(-96 L_0^r
+32 L_1^r+32 L_2^r-16 L_3^r-32 L_4^r+8 L_5^r+32 L_6^r-16 L_8^r-\frac{5}{576 \pi^{2}}\biggr) \nonumber\\
&+& \frac{8 M_{\pi }^3 M_K L_5^r}{F_{\pi }^4} +\frac{\mu _{\pi }}{F_{\pi }^2 \left(M_K^2-M_{\pi }^2\right)}
\biggl(-\frac{M_K^4}{4}-\frac{1}{2} M_{\pi } M_K^3+\frac{13}{4} M_{\pi }^2 M_K^2+\frac{5}{2} M_{\pi }^3 M_K\biggr)
\nonumber\\
&+& \frac{\mu _K}{F_{\pi }^2}\biggl(-\frac{M_K^3}{M_K-M_{\pi }}-\frac{M_{\pi } M_K^2}{M_K-M_{\pi }}
-\frac{4 M_{\pi }^2 M_K^2}{M_K^2-M_{\pi }^2}-\frac{M_{\pi }^2 M_K}{2 \left(M_K-M_{\pi }\right)}
+\frac{M_{\pi }^3}{2 \left(M_K-M_{\pi }\right)}\biggr)\nonumber\\
&+& \frac{\mu _{\eta }}{F_{\pi }^2 \left(M_K^2-M_{\pi }^2\right)}\biggl(-\frac{4 M_{\pi }^2 M_K^4}{9 M_{\eta }^2}
-\frac{M_{\pi }^4 M_K^2}{18 M_{\eta }^2}+\frac{3 M_K^4}{4}+\frac{3}{2} M_{\pi } M_K^3+\frac{11}{4} M_{\pi }^2 M_K^2
-\frac{3}{2} M_{\pi }^3 M_K\biggr)\nonumber\\
&+& \frac{M_{\pi }^2 M_K^2 \bar{J}_{\pi K}(s_0)}{F_{\pi }^4}
+ \frac{\bar{J}_{\pi K}(u_0)}{F_{\pi }^4}\biggl(\frac{M_K^4}{8}+\frac{1}{2} M_{\pi } M_K^3
-\frac{1}{4} M_{\pi }^2 M_K^2+\frac{1}{2} M_{\pi }^3 M_K+\frac{M_{\pi }^4}{8}\biggr) \nonumber\\
&+& \frac{\bar{J}_{K\eta}(u_0)}{F_{\pi }^4}\biggl(\frac{M_K^4}{72}+\frac{1}{18} M_{\pi } M_K^3
-\frac{7}{12} M_{\pi }^2 M_K^2+\frac{1}{18}M_\pi^3M_K+\frac{M_{\pi }^4}{72}\biggr)\nonumber\\
&+& \frac{\bar{\bar{J}}_{\pi K}\left(u_0\right)}{F_{\pi }^4}\biggl(-\frac{M_K^4}{8}-\frac{1}{2} M_{\pi } M_K^3
-\frac{3}{4} M_{\pi }^2 M_K^2-\frac{1}{2} M_{\pi }^3 M_K-\frac{M_{\pi }^4}{8}\biggr) \nonumber\\
&+& \frac{\bar{\bar{J}}_{K\eta}\left(u_0\right)}{F_{\pi }^4}\biggl(-\frac{M_K^4}{72}-\frac{1}{18} M_{\pi } M_K^3
-\frac{1}{12} M_{\pi }^2 M_K^2-\frac{1}{18} M_{\pi }^3 M_K-\frac{M_{\pi }^4}{72}\biggr)
-\frac{5 M_K^4}{384 \pi ^2 F_{\pi }^4} -\frac{5 M_{\pi } M_K^3}{192 \pi ^2 F_{\pi }^4}\nonumber\\
&-&\frac{M_{\pi }^3 M_K}{192 \pi^2 F_{\pi }^4}+\frac{M_{\pi } M_K}{F_{\pi }^2}-\frac{M_{\pi }^4}{384 \pi ^2 F_{\pi }^4}
\biggr]~,\label{eq:a0beta}
\end{eqnarray}
where  $\mu_P = (M_P^2/32\pi^2F_\pi^2)\ln(M_P^2/\mu^2)$, with $\mu$ the renormalization scale,
and the functions $\bar{J}_{PQ}(s), \bar{\bar{J}}_{PQ}(s)$
are given in Ref.~\cite{Guruswamy:2020uif}.
The $\{L_i^r\}$ are the $\mathcal{O}(p^4)$ LECs in SU$(4|1)$ PQChPT, among which
$L_{1-8}^r$ are numerically identical with those in the ordinary three-flavor ChPT.
Only $L_0^r$ is new, and can be readily solved from the equation above.

\begin{table}
\begin{center}
\begin{tabular}{|c|c|c|c|c|}
	\hline
	\rule[-1ex]{0pt}{2.5ex} Ensemble & $aM_{\pi}$& $aM_{K}$ & $aM_{\eta}$ & $aF_\pi$ \\
	\hline
	\rule[-1ex]{0pt}{2.5ex} A30.32 & 0.1239(3) & 0.236(7) & 0.314(17) &0.06452(21)  \\
	\hline
	\rule[-1ex]{0pt}{2.5ex} A40.24 & 0.1452(5)  & 0.241(7)
	  & 0.317(7) & 0.06577(24)
	    \\
	\hline
\end{tabular}
\caption{Basic parameters of the two lattice ensembles in Ref.\cite{Helmes:2018nug}, with
  lattice spacing $a=0.0885(36)$~fm. The uncertainties are mainly from finite-volume effects.}
\label{messonmasses}
\end{center}
\end{table}

\begin{table}
	\begin{center}
		\begin{tabular}{|c|c|c|}
			\hline
		\multicolumn{3}{|c|}{$\mu_{\pi K}a_{0}^{\beta}$}  \\
			\hline
		Ensemble	& \textbf{E1} & \textbf{E2}   \\
			\hline
		A30.32	& $-0.0956(91)$ & $-0.0961(84)$   \\
			\hline
		A40.24	&  $-0.1152(144)$ &  $-0.1142(131)$      \\
			\hline
		\end{tabular}
	        \caption{The unphysical scattering length $a_0^\beta$
                  evaluated from the ensembles A30.32 and A40.24. The uncertainties are mainly
                  from thermal pollutions, which are treated using two different methods (\textbf{E1}
                  and \textbf{E2}).} \label{Scattlength}
	\end{center}
\end{table}

\section{Extraction of $L_{0}^{r}$}

The unphysical scattering length $a_0^\beta$ is an essential input in our study. It is obtained from
the analysis of the lattice-volume-dependence of the discrete energy levels corresponding to
the combination $T_a-T_b$ through the standard L\"{u}scher formula~\cite{Luscher:1986pf}. In this
work, we utilize the results in 
Ref.~\cite{Helmes:2018nug} for the $K\pi$ system in the $I=3/2$ channel where the correlation
functions $C_a(\tau)$ and $C_b(\tau)$ that correspond to the two contractions in Fig.\ref{fig:TaTb}
were separately calculated. The calculation was based on
the $N_f = 2+1+1$ twisted mass lattice QCD. For our analysis we have considered the ensembles
A30.32 and A40.24 for the determination of the discrete ground-state energies of the $K\pi$
system. The basic parameters of each ensemble are summarized in Table~\ref{messonmasses}.
	
Only results that correspond to the physical combination $C_a(\tau)+C_b(\tau)$ were displayed in
Ref.~\cite{Helmes:2018nug} for obvious reasons. In this project, we acquired the unphysical
S-wave scattering length $a_0^\beta$ directly from the authors of that paper, who obtained its
value through an unpublished analysis of the volume-dependence of the discrete energy levels
extracted from the unphysical combination $C_\beta(\tau)=C_a(\tau)-C_b(\tau)$ \cite{FP}:
First, the energy shift $\delta E_\beta=E_\beta^{\pi K}-M_\pi-M_K$ was obtained as a function
of the lattice size $L$ from the exponential behavior of $C_\beta(\tau)$ at large Euclidean
time $\tau$. The scattering length $a_0^\beta$ at infinite volume was then  computed using the
single-channel, Taylor-expanded L\"{u}scher formula
\begin{equation}
\delta E_\beta=-\frac{2\pi a_0^\beta}{\mu_{\pi K}L^3}\left(1+c_1\frac{a_0^\beta}{L}
+c_2\frac{(a_0^\beta)^2}{L^2}\right)+\mathcal{O}(L^{-6})
\end{equation}
as in Eq.(14) of Ref.~\cite{Helmes:2018nug}. Here, $\mu_{\pi K}$ is the reduced mass of the $K\pi$
system and $c_{1,2}$ are known coefficients. 
The final outcomes are summarized in Table~\ref{Scattlength}. The main uncertainty of $a_0^\beta$
is systematic, which comes from the unwanted time-dependent contributions at finite $\tau$
(i.e. the ``thermal pollutions''). Their effects were studied using two different methods
labeled as \textbf{E1} (weighting and shifting) and \textbf{E2} (dividing out the pollution),
respectively~\cite{Dudek:2012gj}. 

To solve for $L_0^r$ using Eq.~\eqref{eq:a0beta},  we further require the values of
all the physical LECs. We took their values at $\mu=M_\rho=770$~MeV from Ref. \cite{Bijnens:2014lea}:
\begin{eqnarray}
10^3L_{1}^{r} &=& 1.11(10)~, ~~ 10^3L_{2}^{r} = 1.05(17)~,~~ 10^3L_{3}^{r} = -3.82(30)~,~~
10^3L_{4}^{r}= 1.87(53)~,\nonumber\\
10^3L_{5}^{r} &=& 1.22(06)~, ~~ 10^3L_{6}^{r} = 1.46(46)~,~~ 10^3L_{8}^{r}= 0.65(07)~. 
\end{eqnarray}  
With all the above, we may now compute $L_0^r$ straightforwardly. The outcome at $\mu=M_\rho$ reads:
\begin{eqnarray}
{\rm A30.32~,~Method~ E1} &:& 10^3L_{ 0}^{r} = 0.78(21)(25)(7)(7)(2)\nonumber\\
{\rm A30.32~,~Method~ E2} &:& 10^3L_{ 0}^{r} = 0.77(20)(25)(7)(7)(2)\nonumber\\
{\rm A40.24~,~Method~ E1} &:& 10^3L_{ 0}^{r} = 0.86(25)(25)(7)(4)(2)\nonumber\\
{\rm A40.24~,~Method~ E2} &:& 10^3L_{ 0}^{r} = 0.87(22)(25)(7)(4)(2)
\end{eqnarray}  
where the uncertainties come from $a_0^\beta$, the physical LECs, the higher-order ChPT corrections,
the (lattice) meson masses and $F_\pi$, respectively. In particular, the higher-order ChPT
corrections are estimated by multiplying the central value with the usual chiral suppression
factor $M_K^2/(4\pi F_\pi)^2$. We see that the values of $L_0^r$ from all four
determinations are consistent with each other within the error bars, so we may simply
quote the number with the smallest theory uncertainty, namely the one from the ensemble A30.32
with method \textbf{E2}:
\begin{equation}
\label{finalres}
L_0^r(M_\rho)=0.77(33)\cdot 10^{-3}~.    
\end{equation}

Finally, we comment on the relation between this result and the corresponding LEC in SU$(4|2)$.
In principle, relations between LECs in the two-flavor and three-flavor ChPT can be obtained
by integrating out the strange quark in the latter. Possible PQChPT extensions of an
ordinary two-flavor ChPT are SU$(3|1)$, SU$(4|2)$, SU$(5|3)$..., etc, but only SU$(4|2)$
onwards possess an $L_0$-dependence at tree-level as it requires at least four
fermionic quarks. Similarly, possible PQChPT extensions of a three-flavor ChPT are SU$(4|1)$,
SU$(5|2)$, SU$(6|3)$...\, . In Ref.~\cite{Guruswamy:2020uif} we chose the simplest version
which is SU$(4|1)$. After integrating the strange quark, it reduces to SU$(3|1)$ that does
not depend on $L_0$ at tree-level. Therefore, it is not possible to discuss the matching
between $L_0$ in two- and three-flavors based on the theory setup in
Ref.~\cite{Guruswamy:2020uif}. For that, one would have to repeat the calculations 
using a larger graded algebra, such as SU$(5|2)$. This is of great interest because it may provide
new insights to the apparent disagreement between the determination of $L_0^r$ at SU(4$|$2)
from NLO and NNLO. 
However, it goes beyond the scope of this letter and will be carried out in follow-up studies.

\section{Summary}

In this work, we have for the first time determined the unphysical LEC $L_0^r(M_\rho)$ in the
simplest PQ-extension of the three-flavor ChPT through an NLO analysis of contraction diagrams
in $K\pi$ scattering, $L_0^r = 0.77(33) \cdot 10^{-3}$. Utilizing the precise data from
the ETM collaboration for $K\pi$ scattering in the $I=3/2$ channel, we
control the absolute uncertainty in this LEC to $3.3\times 10^{-4}$, which is better than the previous
determinations of the corresponding LEC for two flavors
in Ref.~\cite{Boyle:2015exm} that made use of an NNLO fitting, and in the NLO fitting to
the $\pi\pi$ scattering amplitudes in Refs.~\cite{Acharya:2017zje,Acharya:2019meo} that depends
on more unknown LECs. The major sources of uncertainty in this study are the systematic errors
in the lattice extraction of the unphysical scattering length $a_0^\beta$, as well as the physical
LECs $L_{1-8}^r$. Our work completes the PQChPT Lagrangian at $\mathcal{O}(p^4)$ and prepares
it for the future applications in studies of interesting hadronic observables involving
strange quarks, in synergy with lattice QCD.

\section*{Acknowledgements}

We are very grateful to Ferenc Pitler for making the ETM collaboration data available to us and
for his detailed explanations concerning these. We thank Hans Bijnens for a useful communication.
This work is supported in part by  the DFG (Projektnummer 196253076 - TRR 110)
and the NSFC (Grant No. 11621131001) through the funds provided
to the Sino-German CRC 110 ``Symmetries and the Emergence of
Structure in QCD", by the Alexander von Humboldt Foundation through the Humboldt
Research Fellowship, by the Chinese Academy of Sciences (CAS) through a President's
International Fellowship Initiative (PIFI) (Grant No. 2018DM0034), by the VolkswagenStiftung
(Grant No. 93562), and by the EU Horizon 2020 research and innovation programme, STRONG-2020 project
under grant agreement No. 824093.

\end{document}